\documentclass[letterpaper]{article}

\usepackage{natbib,alifeconf}  %% The order is important
\usepackage{url,hyperref,cleveref}
\usepackage{booktabs}
\usepackage{subcaption}
\usepackage{float}
\title{Emergent time-keeping mechanisms in a deep reinforcement learning agent performing an interval timing task}

\author{
    Amrapali Pednekar $^{1}$,
    Alvaro Garrido$^{1}$,
    Pieter Simoens$^{1}$, \and 
    Yara Khaluf$^{1,2}$ \\
    \mbox{}\\ 
    $^1$ IDLab, Department of Information Technology, Ghent University - imec, Belgium (Email: amrapali.pednekar@ugent.be) \\
    $^2$Department of Social Sciences, Wageningen University and Research, 6706KN, Wageningen, The Netherlands \\
} % email of corresponding author

\begin{document}

\maketitle

\begin{abstract}
    % Abstract length should not exceed 250 words
    Drawing parallels between Deep Artificial Neural Networks (DNNs) and biological systems can aid in understanding complex biological mechanisms that are difficult to disentangle. Temporal processing, an extensively researched topic, is one such example that lacks a coherent understanding of its underlying mechanisms. In this study, we investigate temporal processing in a Deep Reinforcement Learning (DRL) agent performing an interval timing task and explore potential biological counterparts to its emergent behavior. The agent was successfully trained to perform a duration production task, which involved marking successive occurrences of a target interval while viewing a video sequence. Analysis of the agent’s internal states revealed oscillatory neural activations, a ubiquitous pattern in biological systems. Interestingly, the agent’s actions were predominantly influenced by neurons exhibiting these oscillations with high amplitudes and frequencies corresponding to the target interval. Parallels are drawn between the agent’s time-keeping strategy and the Striatal Beat Frequency (SBF) model, a biologically plausible model of interval timing. Furthermore, the agent maintained its oscillatory representations and task performance when tested on different video sequences (including a blank video). Thus, once learned, the agent internalized its time-keeping mechanism and showed minimal reliance on its environment to perform the timing task. A hypothesis about the resemblance between this emergent behavior and certain aspects of the evolution of biological processes like circadian rhythms, has been discussed. This study aims to contribute to recent research efforts of utilizing DNNs to understand biological systems, with a particular emphasis on temporal processing. 
\end{abstract}

% Choose one of: Full Paper, Summaries, or Late Breaking Abstracts 
% Submission type: \textbf{Full Paper}\\
Code available at: \url{https://github.com/decide-ugent/drl-video-timer}

\section{Introduction}
A recent research interest has emerged in drawing parallels between Deep Artificial Neural Networks (DNNs) and biological systems \citep{kanwisher2023using, richards2019deep}. The similarities between the two are noteworthy because, DNNs exhibit emergent behavior and are not explicitly engineered to mimic biological systems. Unlike biological systems, which are a result of complex evolutionary adaptations, DNNs rely on relatively simple optimization techniques and architectures to solve problems. These two distinct systems converging to similar behaviors suggests important implications about the nature of the problem being solved. It may indicate that certain behaviors and representations are fundamentally useful to find a solution, and that problems presumed to require complex solutions might, in fact, be addressed through simpler means. Additionally, DNNs, with their flexibility for ablation studies and architectural manipulations, provide a powerful platform for systematically testing different hypotheses. Hence, identifying similarities between the two systems can facilitate better understanding of biological systems. 

Temporal processing is an extensively researched topic in both psychology and neuroscience \citep{wittmann2009inner}. However, the lack of a dedicated sensory organ for perceiving time, and scattered evidence regarding the involvement of multiple brain regions, makes it difficult to form a coherent understanding of temporal processing in humans and animals. Interval timing provides a framework to study these temporal processing mechanisms. The ability of humans and animals to successfully produce, reproduce, or categorize different temporal durations suggests the existence of an internal time-keeping mechanism. Recent research has begun exploring interval timing also in deep reinforcement learning (DRL) agents \citep{deverett2019interval, lin2023temporal}. These studies additionally explore potential biological counterparts to the internal time-keeping mechanisms that emerge in the DRL agent performing interval timing tasks. They put forth a discussion about using advancements in Artificial Intelligence (AI) to enhance our understanding of temporal encodings in the brain.

In this study, we investigate the internal time-keeping mechanisms of a DRL agent trained on a duration production task \citep{thones2019standard}. We show that the agent develops a robust internal time-keeping mechanism, characterized by oscillatory activation patterns, a motif commonly observed in biological systems.  Furthermore, the agent's timing strategy shares high-level functional similarities with the Striatal Beat Frequency (SBF) model, a biologically plausible model of interval timing \citep{matell2000neuropsychological}. Finally, the robust and internalized time-keeping mechanism developed in the agent is compared to certain aspects of the evolution of biological processes like circadian rhythms.  Through this study, we aim to contribute to research that draws parallels between temporal processing in artificial agents and biological systems. It is our view that such efforts may provide alternative approaches to understanding temporal processing in humans and animals.

\section{Methodology}

\begin{figure}
    \centering
    \includegraphics[width=175px]{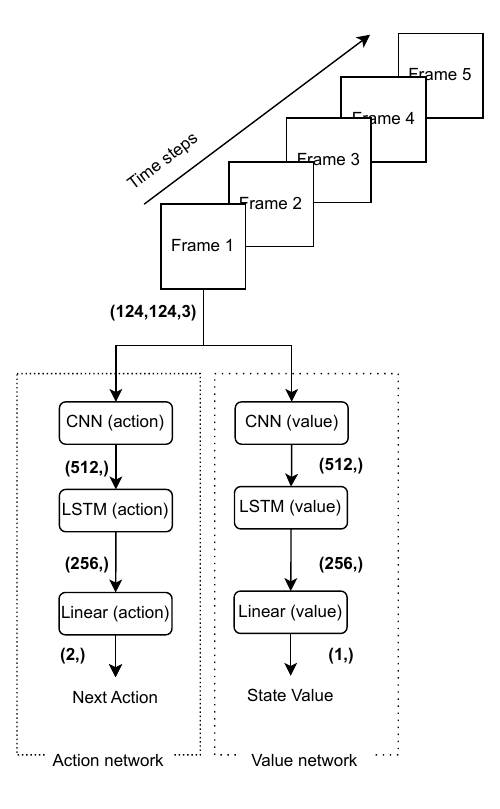}
    \caption{Overview of the DRL agent architecture and input processing pipeline: The DRL agent's architecture consists of an action network (left) and a value network (right), both processing the input frames in parallel. The agent receives one frame at each time step from the environment. The input shape transitions from the original frame to the action and state value, are shown in parentheses. At each time step, the agent can take one of the two actions: “GO” or “Interval,” to advance to the next frame and receive a reward from the environment. 
    }
    \label{method}
\end{figure}

The interval timing task used in this study is a time production task that involved watching a video sequence and successively marking a fixed time interval. The DRL agent received one frame at a time from its environment. Thus, the agent's state (observation) at a given point was the current frame it was viewing. These frames were a part of a video sequence that was a randomly chosen sequence of consecutive frames extracted from a single video. The length of the video sequence was five times the target duration. Thus, for a target duration of 4 time steps, the video sequence comprised of twenty frames and consequently one episode comprised of twenty time steps. The agent had two possible actions: "Go", to proceed to the next video frame and "Interval", to mark an interval (i.e., produce the target interval) and proceed to the next frame.  When the agent produced the correct interval, it received a reward of +1; for incorrect productions or missed intervals, it received a penalty of -1; all other cases generated a reward of 0. Different agents were trained to produce target intervals of two, three, four and five time steps. In the specific examples used in this study, the agent was trained to produce four time steps. Thus, the agent had to select the action "Interval" on every fourth frame. The important findings discussed in this study are consistent across agents trained for other target intervals (see Appendix for details). 

The DRL agent architecture consisted of three primary components: a Convolutional Neural Network (CNN) to extract spatial features from each video frame, a Long Short-Term Memory (LSTM) network to process temporal dependencies across frames, and a linear layer to process the resulting representation and to compute the state value and next action (see \Cref{method}). The agent was implemented using the recurrent policy variant of the Proximal Policy Optimization (PPO) algorithm provided by the Stable-Baselines3 (SB3) library \citep{stable-baselines3}. The reinforcement learning environment was a custom built environment developed using the Gymnasium framework \citep{towers2024gymnasium}. It was designed to deliver sequential video frames and handle action and rewards specific to the target interval time. The analysis conducted in this study focuses exclusively on the action network.

The agent was evaluated on a set of video sequences extracted from videos other than the training video. These test videos were categorized based on their Mean Absolute Frame Difference (MAFD), computed as described in \cite{salih2020dynamic} and averaged over all frames for each video sequence (see Table~1). We use the average MAFD as a simplified metric to quantify the amount of visual information in a video sequence. A blank video, or one in which the same frame is repeated throughout (i.e., constant frame video),  yields a MAFD of zero and thus contains minimal visual information. In contrast, videos with changing frames exhibit higher MAFD values and thus contain more visual information. It is worth noting that, a video sequence composed of random frames exhibits even higher MAFD values, despite lacking meaningful visual information. Thus, although high MAFD reflects greater pixel-level changes between video frames, it does not necessarily correspond to meaningful visual content. However, since the agent’s primary objective is not to process visual information as such, but to produce specific intervals, we assume that higher MAFD corresponds to greater visual information from the agent’s perspective, with the caveat that this metric captures only low-level variability. 

In order to evaluate the generalizability of the results discussed in this study, different agents were trained using varying input configurations. These can be categorized into three types. First, as mentioned above, agents were separately trained to produce target intervals of two, three, four and five time steps each. This helped ensure consistency of the findings across different temporal durations. Second, different agents were trained on videos with extreme MAFD values, specifically, a blank video, a constant frame video and a random frame video. This helped assess the extent to which the agent relied on its input to perform the time production task and the input video characteristics that may influence the agent’s performance. Lastly, an agent was trained to perform a delayed timing task, in which a cue frame (a red frame) signaled the start of the time production task. Before the cue frame, the agent received a reward of zero regardless of its actions. After the cue frame, the agent received rewards as described above for performing the time production task. The position of the cue frame was randomly selected to occur no earlier than one-third of the video sequence length and no later than a point that allowed at least four target durations to be produced before the end of the episode. The episode length was also increased to ten times the target duration (instead of five). These design choices ensured that there were sufficient time steps before and during the time production task to reliably observe neural activations. This variation enabled the identification of neural activation patterns specific to the time production phase and highlighted differences in neural activity before and during the timing task.

\section{Neural oscillations - an emergent time-keeping mechanism in the artificial agent}

The agent successfully managed to learn the timing task and was able to produce the target interval consecutively throughout the video sequence. Analysis of the agent’s internal states revealed that the LSTM layer played a crucial role in the development of an intrinsic timekeeping mechanism. This was expected since the LSTM network, which is a type of artificial Recurrent Neural Network (RNN), is designed to extract temporal features from sequential data \citep{hochreiter1997long, yu2019review}. An analysis of the LSTM layer's hidden state activations reveals emergent high amplitude oscillatory patterns with frequencies correspond to the target interval.

\begin{figure}[ht]
    \centering
    \includegraphics[width=\linewidth]{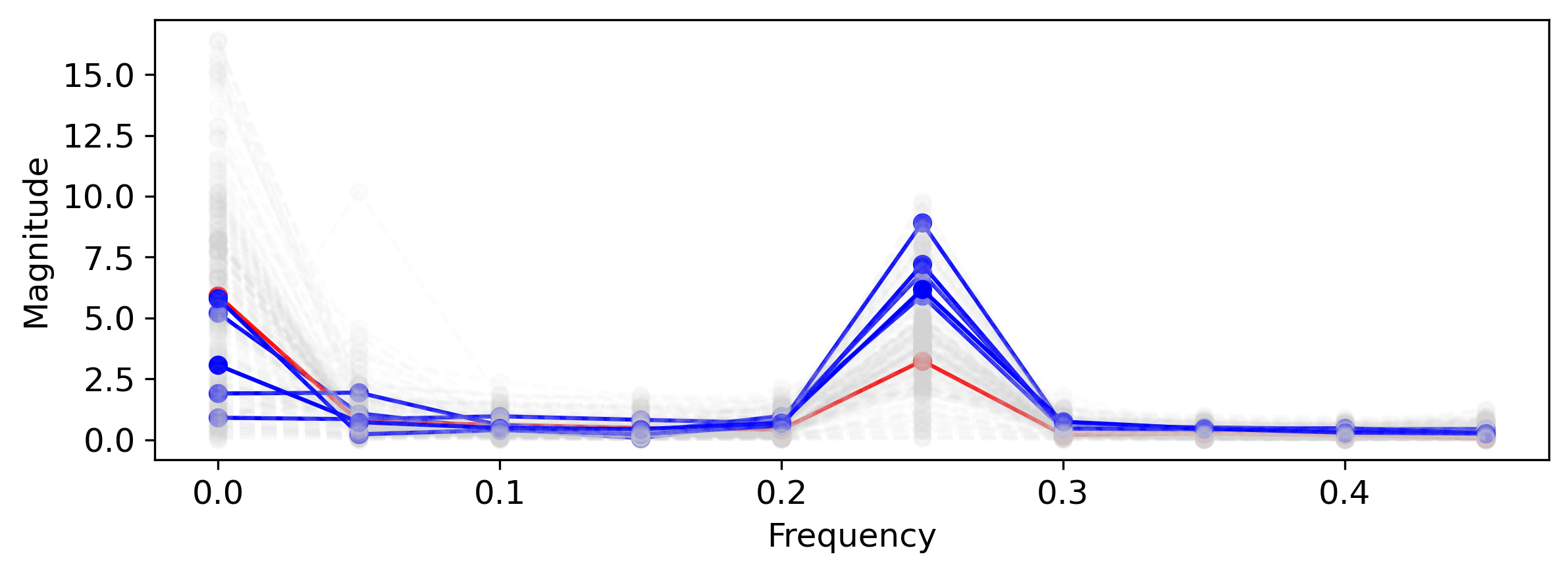}
    \caption{Fast Fourier Transform (FFT) of the LSTM hidden state activations across time for the time production task with a target duration of four time steps. The blue lines represent neurons with high weight magnitudes (z-score $>$ 2) for the ``Interval" action in the action network , while the red lines represents neurons with low weight magnitudes (z-score $<$ -2) for both actions. The gray lines correspond to all other neurons. All neurons exhibit oscillations with a frequency of 0.25 (1 peak every 4 time step). 
    }
    \label{fft_hidden_state}
\end{figure}

A Fourier Transform of LSTM hidden state activations revealed a clear oscillatory pattern with frequencies corresponding to the target interval. \Cref{fft_hidden_state} shows the Fast Fourier Transform (FFT) of the hidden state activations across time for a target duration of four time steps. It shows that most neurons exhibit oscillations at a frequency of 0.25, corresponding to one peak every four time steps. Notably, neurons with high weight magnitudes (z-score $>$ 2) for the ``Interval" action in the action network (highlighted in blue in \Cref{fft_hidden_state}) show higher oscillation amplitudes compared to those with lower weight magnitudes (z-score $<$ -2) for both actions (highlighted in red in \Cref{fft_hidden_state}). These results are consistent across different target durations (see Appendix \Cref{fig:appendix_fft} for details). This suggests that the agent's decision to select the `Interval" action is more strongly influenced by neurons exhibiting high amplitude oscillations with frequencies corresponding to the target interval.  

\begin{figure}[ht]
    \centering
    \includegraphics[width=260px]{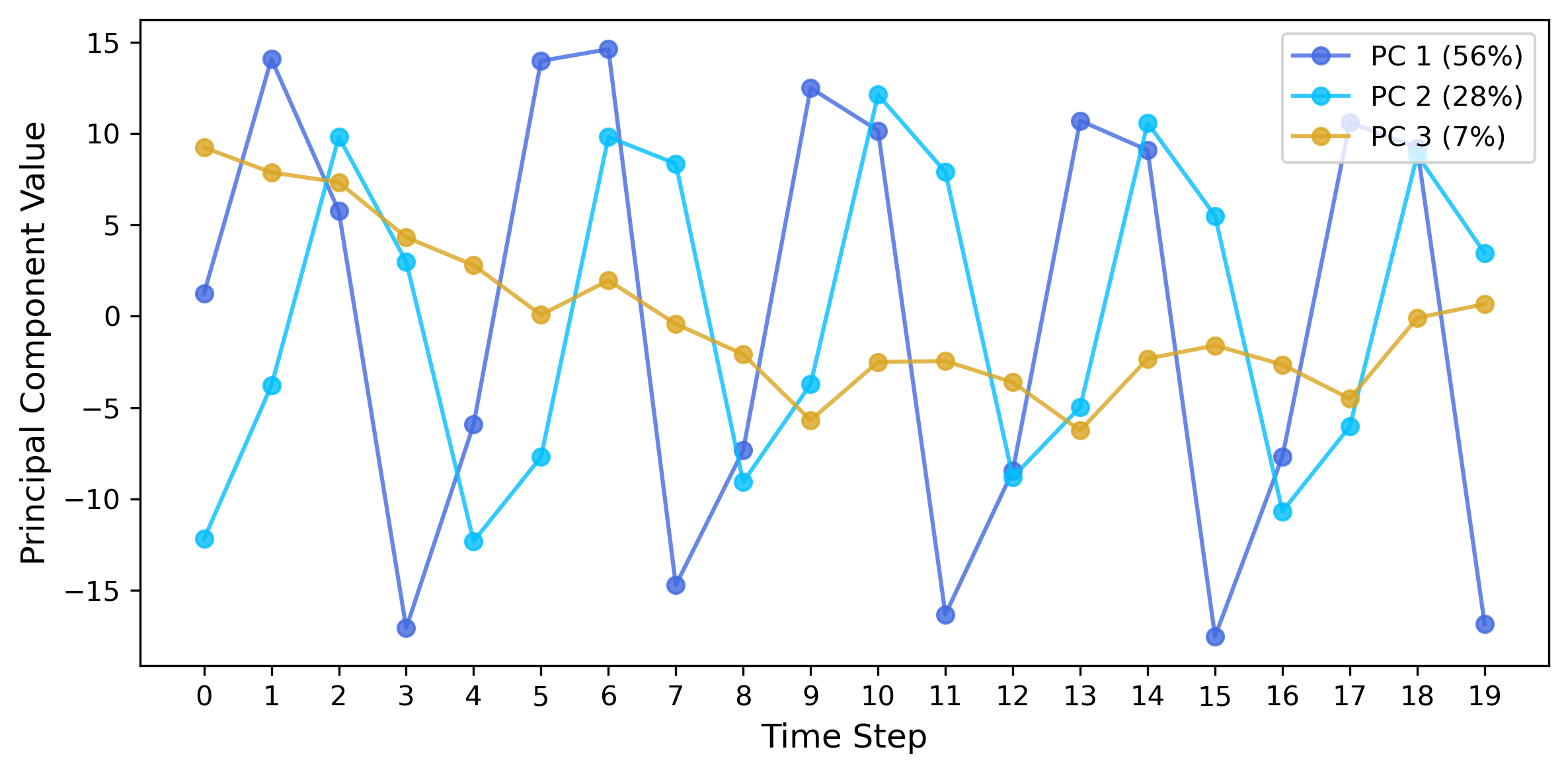}
    \caption{Principal component analysis (PCA) of neural activations of the 256 hidden state neurons in the LSTM network across time for the time production task with a target interval of four time steps. The first and second principal components (in dark and light blue respectively) which exhibit an oscillatory pattern across time steps, explain 56\% and 28\% of variance respectively. The first principal component exhibits an oscillatory pattern with a period matching the target interval (i.e., four time steps)
    }
    \label{pca}
\end{figure}

A Principal Component Analysis (PCA) was conducted to identify dominant patterns in the activations of the 256 hidden state neurons in the LSTM network across time steps. \Cref{pca} shows the results of PCA for a target duration of four time steps. The first two principal components that explained 86\% of variance exhibit oscillations. Interestingly, the first principal component that explains 56\% of variance exhibits oscillations with a period equal to the target interval (i.e., four time steps). The PCA conducted on other durations, namely two, three, and five, also showed similar results where the first principal components that explained more than 50\% of variance exhibited oscillations with a period equal to the target duration (see Appendix \Cref{fig:appendix_pca} for details). This analysis further verified the occurrence of oscillatory patterns across time in the hidden state activations of the DRL agent's LSTM layer, suggesting their potential role in time-keeping due to their phase alignment with the target interval.

Interestingly, a prior study involving a DRL agent trained for time reproduction (with spatial input) did not report such oscillatory patterns in the PCA of the LSTM cell states \citep{deverett2019interval}. The authors concluded that the cell state units in the LSTM exhibit counter-like behavior. However, in \citealp{lin2023temporal}, for certain conditions, oscillating activity has been reported in the RNN units of a DRL agent (with non-spatial inputs) for a duration comparison and an episodic working memory task. Additionally, oscillatory patterns have been observed in a continuous time recurrent neural networks (CTRNNs) optimized using genetic algorithm to perform multiple timing tasks \cite{maniadakis2015integrated}. However, unlike the LSTM or RNN layer of the DRL agents, CTRNNs are known for their biologically plausible interpretation \cite{yu2017continuous}, \cite{beer1995dynamics}. Thus, findings from current and prior studies suggests that oscillatory dynamics may serve as a fundamental mechanism of time-keeping for specific types of interval timing tasks in artificial agents, regardless of the biological plausibility of their architectures. 

Furthermore, for the delayed timing task (see Methodology for details), the PCA and FFT of the LSTM hidden state activations show that high amplitude oscillating patterns emerge predominantly during the time production phase (i.e., after the cue frame onset). \Cref{fig:pca_delayed_time} shows the PCA of LSTM hidden state activations across time for the delayed timing task with a target interval of four time steps. A clear difference in dominant neural activation patterns is observed before and after the cue frame onset, which marks the beginning of the time production task. Moreover, \Cref{fig:pca_fft_delayed_task}a and \Cref{fig:pca_fft_delayed_task}b respectively show the FFT of the LSTM hidden state activations before and after  cue frame onset for the delayed timing task with a target interval of four time steps. High-amplitude oscillations with frequencies approximately around the target interval emerge only after the start of the time production task, that is, after the cue frame onset (as shown in \Cref{fig:pca_fft_delayed_task}b). In contrast, prior to the cue frame onset (\Cref{fig:pca_fft_delayed_task}a), the same neurons, such as those shown in blue in \Cref{fig:pca_fft_delayed_task}a and \Cref{fig:pca_fft_delayed_task}b (which also correspond to neurons with high weights in the action network), exhibit lower amplitude oscillations at frequencies different from the target intervals. This behavior was consistent across different runs with varying cue frame onset positions. This result provides additional evidence that the high amplitude oscillation patterns  with frequencies corresponding to the target interval emerge specifically in response to the time production task. It further supports their role as a potential mechanism for time-keeping in the agent.  

Emergence of oscillatory patterns in the DRL agent's network is noteworthy  because oscillations are a known characteristic of various types of biological neurons \citep{heltberg2021tale, buzsaki2013scaling}, including those involved in temporal processing. Importantly, the DRL agent architecture was not designed to be biologically plausible and it was not provided with any external timing cues or explicitly encoded neural dynamics. Despite this, it successfully learned to perform the timing task, and, remarkably, the hidden state activations of the network converged to an oscillatory pattern with frequencies corresponding to the target interval. These findings suggest that, despite its non-biological architecture, the DRL agent converges to use biologically plausible mechanisms, specifically, oscillatory dynamics, for time-keeping in a dedicated time production task.

\section{Parallels with a biologically plausible timing model}

A biologically plausible model of interval timing called the Striatal Beat Frequency model (SBF), describes interval timing in terms of coincidence detection in neural oscillations \cite{matell2000neuropsychological}. According to this model, the striatal spiny neurons function as coincidence detectors of cortical neural oscillations. These coincidence detectors fire when a set amount of coincidental input activity (i.e., coincident oscillations of the cortical neurons) is reached. According to \cite{buhusi2005makes} and \cite{ matell2000neuropsychological}, for tasks involving the production of a learned interval, there is a burst of dopaminergic activity at the trial onset. This triggers synchronization in oscillation of the cortical neurons. Additionally, another burst in activity of dopaminergic neurons occurs at the expected time of reward which corresponds to the target interval. This causes an update in the weights of corticostriatal synapses and consequently triggers the firing of the striatal spiny neurons (or  the coincidental detectors). As a result, the firing rate of these coincidence detectors can encode information about a previously learned interval. Two high-level functional abstractions of the SBF model are considered in this study to explore whether the DRL agent’s emergent time-keeping mechanism resembles the biologically plausible timing model. 

\begin{figure*}[t] 
  \centering

  \includegraphics[width=\textwidth, height=5.5cm, keepaspectratio]{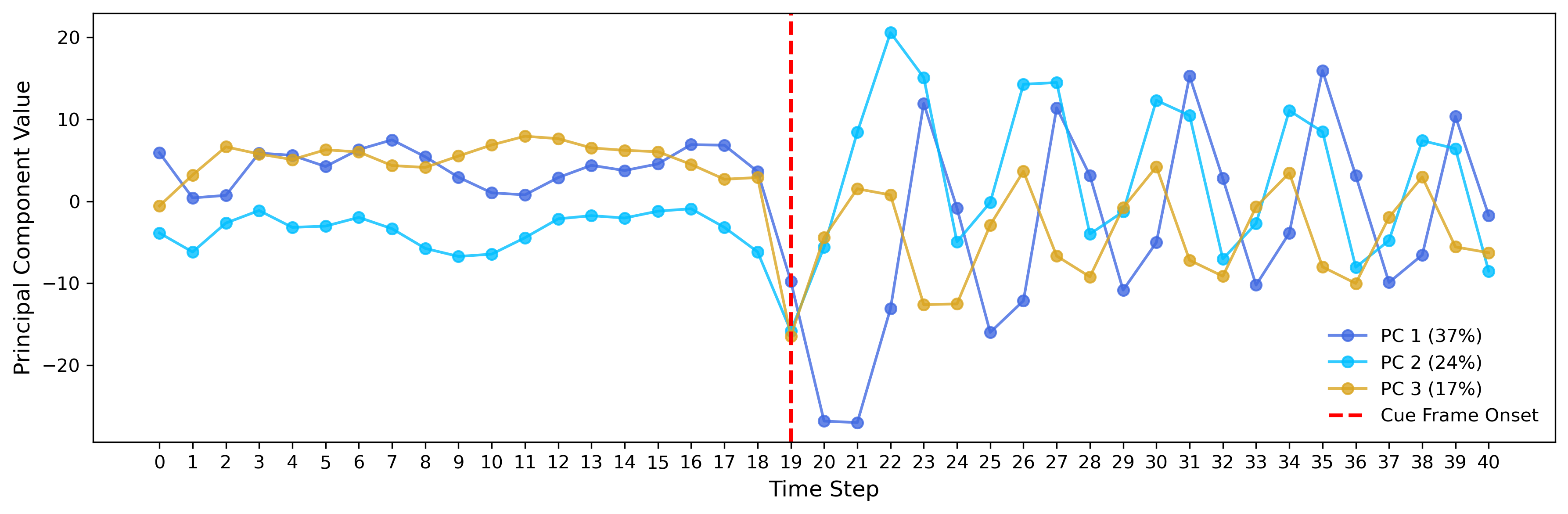}
  \caption{Delayed timing task: Principal Component Analysis (PCA) of the 256 LSTM hidden state activations across time for the delayed timing task with a target interval of four time steps. The first and second principal components explain 37\% and 24\% variability, respectively. A noticeable shift in activation patterns occurs after the cue frame onset, which marks the beginning of the time production phase. During this phase, the first principal component exhibits an oscillatory pattern with a period matching the target interval. }
  \label{fig:pca_delayed_time}

  \vspace{0.5em}  % vertical spacing between rows

  % Two bottom figures side by side
  \begin{subfigure}[b]{0.49\textwidth}
    \includegraphics[width=\linewidth]{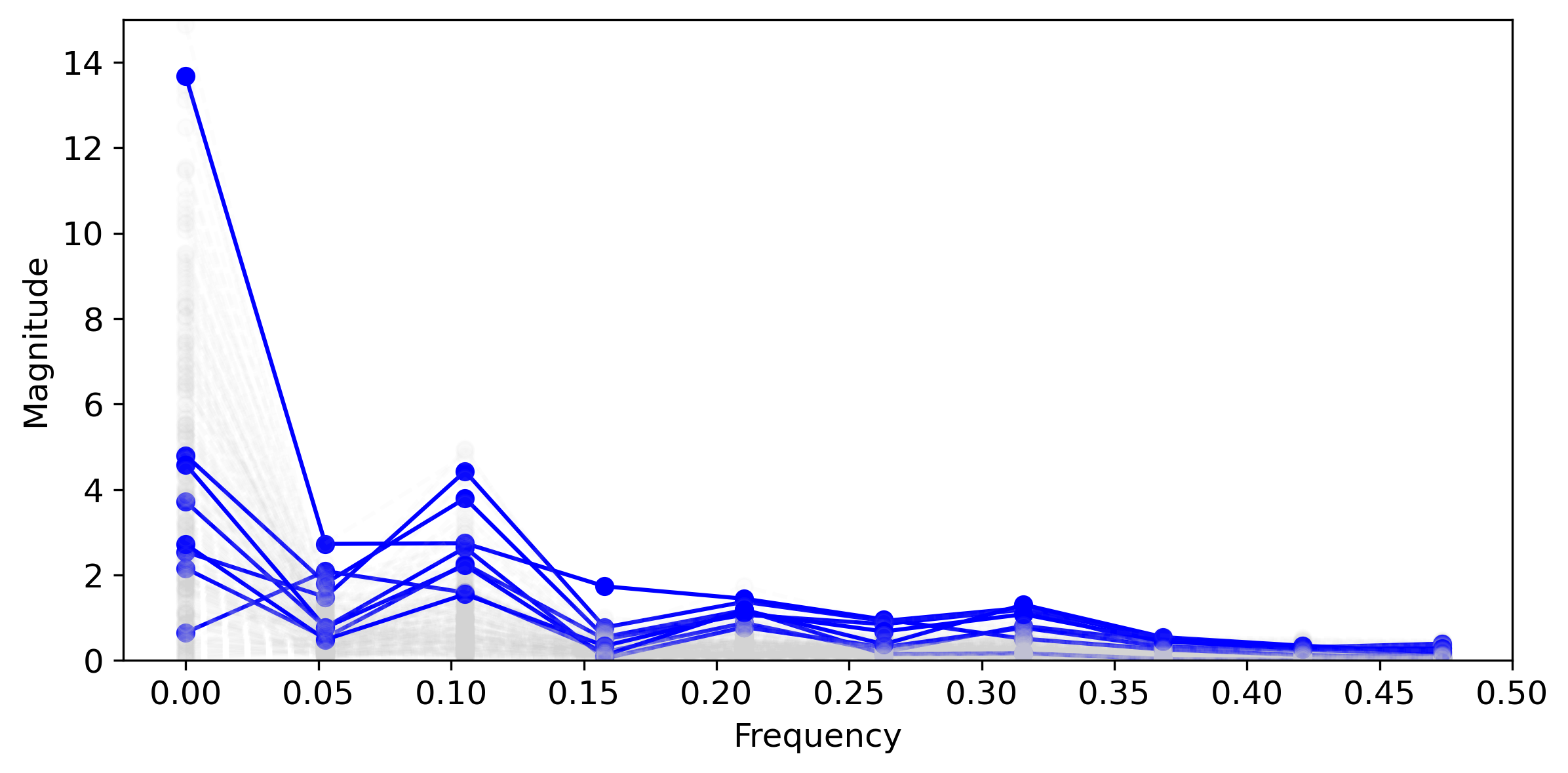}
    \caption{Before cue frame onset}
    \label{fig:fft_1_delayed_time}
  \end{subfigure}
  \hfill
  \begin{subfigure}[b]{0.49\textwidth}
    \includegraphics[width=\linewidth]{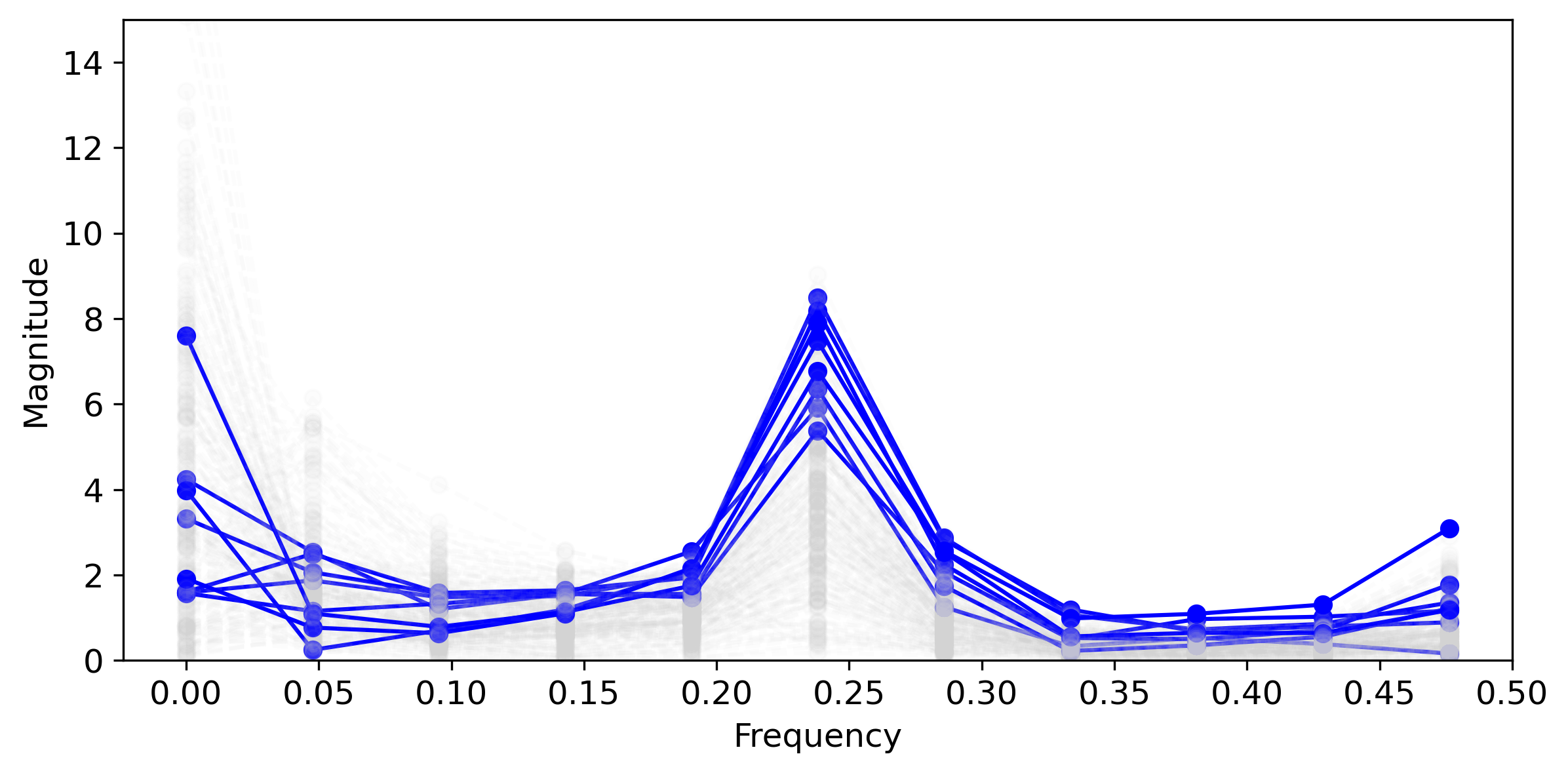}
    \caption{After cue frame onset}
    \label{fig:fft_2_delayed_time}
  \end{subfigure}

  \caption{Delayed timing task: Fast Fourier Transform (FFT) of the LSTM hidden state activations over time for the delayed time production task with a target duration of four time steps. The blue lines represent neurons with high weight magnitudes (z-score $>$ 2) for the `Interval” action in the action network. The gray lines correspond to all other neurons. Before cue frame onset (\Cref{fig:pca_fft_delayed_task}a), the activations exhibit low amplitude oscillations with frequencies different from the target interval. After cue onset (\Cref{fig:pca_fft_delayed_task}b), the time production task begins, and the activations exhibit high amplitude oscillations with frequencies approximately equal to the target interval (i.e., 0.25). }
  \label{fig:pca_fft_delayed_task}
\end{figure*}

First, changes in neural dynamics occur at trial onset due to a burst of dopaminergic activities. To explore the changes in neural dynamics of the DRL agent at trial onset we consider the delayed timing task with a target interval of four time steps (see Methodology for details about this task).  The LSTM hidden state activations of the DRL agent trained to perform this task were divided into two segments: before and after the cue frame onset. The cue frame onset marked the beginning of the time production trial. A FFT was applied to both segments independently to assess the presence and characteristics of oscillatory patterns in each segment. Across multiple runs with varying cue frame positions, the LSTM hidden state activations consistently shifted from lower amplitude oscillations with unrelated frequencies before the cue frame onset to high amplitude oscillations with frequencies corresponding to the target interval after the cue frame onset. Additionally, the PCA (shown in \Cref{fig:pca_delayed_time}), also revealed that at the cue frame onset, there is a change in the dominant activation pattern of the LSTM hidden state activations. Since the only inputs the DRL agent receives is the next frame and the corresponding reward, and since the results are consistent across multiple runs with different cue positions, this change in neural dynamics is likely driven by the reward structure. Thus, the change in neural dynamics of the SBF model due to dopaminergic activities at trial onset can be considered functionally analogous to changes in the DRL agent's neural dynamics likely driven by its reward structure at cue frame onset.

\begin{figure*}[ht]
    \centering
    \includegraphics[width=\linewidth, height=5cm, keepaspectratio]{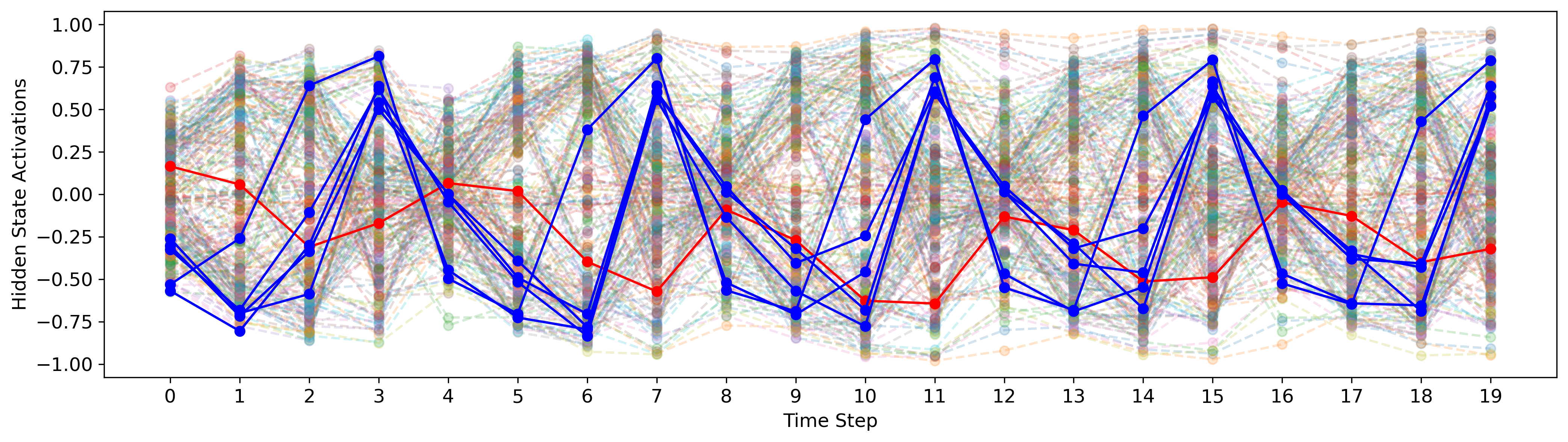}
    \caption{Activations of the 256 LSTM hidden state neurons across time for the time production task with a target duration of 4 time steps. Neurons shown in blue had high weight magnitudes (z-score $>$ 2.0) for the `Interval" action in the agent's action network. While those in red had the least magnitudes (z-score $<$ -2.0) for both actions. Thus, neurons with high amplitude oscillating patterns had the highest contribution in action selection. These neurons peak at the reward generating time steps (i.e., every fourth time steps). 
    }
    \label{raw_activations}
\end{figure*}

Second, these changes in neural dynamics are used by a separate downstream mechanism to produce responses at the target interval. In the DRL agent, the action network serves as the downstream mechanism. It selects an action at each time step based on the weighted sum of the LSTM layer’s neural activations. Thus, the weights assigned to each neuron in the LSTM layer determines the action network’s sensitivity to changes in that neuron's activations. The neurons with high weight magnitudes (z-score $>$ 2) for the `Interval" action are shown in blue, while neurons with low weight magnitudes (z-score $<$ -2) for both actions are shown in red in all the FFT figures. The FFT analysis  (\Cref{fft_hidden_state}, \Cref{fig:pca_fft_delayed_task}a and \Cref{fig:pca_fft_delayed_task}b) reveals that neurons with high weights in the action network for "Interval" action majorly exhibit high-amplitude oscillations with frequencies corresponding to the target interval. Additionally, in the original time production task, the raw activations of these neurons (shown in \Cref{raw_activations}) across different target durations also exhibit a high amplitude oscillatory pattern with peaks at the target interval (see Appendix \Cref{fig:appendix_raw_activations} for raw activations of other target durations). These peaks also align with reward-generating time steps, suggesting that this behavior is likely influenced by the agent’s reward structure. Consequently, at the target interval, these neurons exhibit high activation values, increasing the weighted sum for the "Interval" action. At non-target time steps, their lower activations reduce this weighted sum. Therefore, the action network effectively detects high values of these neurons and triggers the interval action when the weighted sum exceeds a threshold. Functionally, this suggests a possible analogy between the action network and the coincidence detector in the SBF model.

It is important to emphasize that the analogies discussed above are exploratory and are based on high-level functional abstractions. Some important differences in context of the comparisons presented above are as follows. In the SBF model, the trial onset leads to synchronization in frequencies of the cortical neurons without a change in the frequency or amplitude of these neurons. On the other hand, in the DRL agent's time-keeping mechanism, the neurons change their frequency and amplitudes and assume high amplitude oscillations with frequencies corresponding to target interval. Additionally, in the SBF model, the coincident detectors fire when maximum number of neurons peak simultaneously, marking the target interval. In contrast, the DRL agent’s action network selects the `Interval” action when the weighted sum of neural activations is greater than that of the `Go” action.

% align y axis 

\section{Internalized time-keeping in the artificial agent}

To ensure that the agent was not merely capturing visual characteristics of the training video, but had instead learned a general timing mechanism to produce the target interval, we tested it on different types of videos (see Methodology). This evaluation is analogous to determining whether humans are using specific visual cues in a video to mark the target interval, rather than estimating duration based on an internal sense of time. 

Interestingly, the agent was able to successfully perform the time production task even when presented with a blank or constant frame video. Instead, its performance was slightly affected for video sequences with a higher MAFD (see Table~1). Thus, the agent successfully learns a general time-keeping mechanism and does not use visual characteristics of its input video as markers. Furthermore, it internalizes this time-keeping mechanism such that it performs optimally with minimal information from its environment.

The oscillating patterns in the LSTM hidden state activations which are supposed to have emerged as the agent's time-keeping mechanism, were preserved across all test video sequences. The FFT analysis (not shown here) of the hidden state activations from all test video sequences indicated the same oscillation frequencies as in \Cref{fft_hidden_state} but with different magnitudes.
 
In order to check whether the agent relies on visual features of its input during training to learn the time production task, we trained three separate DRL agents (keeping the architecture and training iterations same as the original) with blank video frames, constant video frames and random video frames. It is worth noting that these agents failed to learn the timing task for all three types of videos. This indicates that the agent uses its visual input to extract temporal information during training. The failure of the agent trained on random video frames is especially noteworthy because this indicates that the agent relies on coherent visual (temporal) information and not just pixel-level changes to develop its time-keeping mechanism. Given that time-keeping emerges in the LSTM layer which is known to rely on temporal coherence in its input to perform effectively, the success of the agent on coherent videos, as opposed to an incoherent one seems plausible. Thus, the time-keeping mechanism developed by the agent through interactions with a temporally coherent environment is eventually internalized and is minimally affected by changes in the environment or input coherence.   

A potentially related phenomena can be observed in the evolution of circadian rhythms in biological systems. Circadian clocks in organisms are believed to have developed as a strategy of photosynthesizing organisms to cope with day and night fluctuations from the environment  \cite{hut2011evolution}. These organisms developed oscillating molecular behaviors (typically with periods close to 24 hours) in order to develop different molecular processes corresponding to daylight and darkness. Notably, in some species, these circadian rhythms persist even in the absence of external cues, such as in constant darkness \cite{seki2022evolution}. This highlights how internal mechanisms, initially shaped by environmental interactions in biological systems, can become self-sustaining. We hypothesize that, the DRL agent's behavior related to first learning a task by interacting with its environment, and later internalizing it to be able to perform the task irrespective of the environments input has some parallels with the evolution of biological processes like circadian clocks. 

While the evolution of biological processes such as circadian rhythms involves highly complex mechanisms, the behavior exhibited by the DRL agent emerges from a much simpler setup. Therefore, these processes should not be considered directly analogous. Rather, the aim of this discussion is to propose hypotheses suggesting potential correlations between the behavior of a DRL agent performing a basic timing task and certain features of biological systems. Further evidence is needed to substantiate these correlations and to better understand their implications.

\begin{table}[ht]
\centering
\begin{tabular}{cc}
\toprule
\textbf{Average Frame Difference} & \textbf{Average Reward} \\
\midrule
0.00  & 5.00 \\
3.38  & 4.99 \\
4.85  & 4.96 \\
12.11 & 4.76 \\
4.60  & 4.69 \\
4.35  & 4.69 \\
4.05  & 4.66 \\
5.16  & 4.59 \\
34.7  & 4.35 \\
\bottomrule
\end{tabular}
\caption{Reward across videos with different levels of Mean Absolute Frame Difference (MAFD) \citep{salih2020dynamic} averaged across all frames of the video sequence. The maximum achievable reward per episode is 5. While the agent generally performs well, its performance is slightly affected for video sequences with higher MAFD. Average frame difference of zero corresponds to a blank or constant frame video sequence and the average frame difference of 34.7 corresponds to a random video sequence.}
\end{table}

\section{Discussion}

The DRL agent learned to perform the duration production task and exhibited neural patterns and behaviors that are comparable to those observed in biological systems and biologically plausible models of timing. First, it converged to an internal state with oscillatory neural activations, a ubiquitous pattern in biological neurons. Second, the emergent time-keeping mechanism exhibited high-level functional similarities with the SBF model of timing. Third, the agent initially learned its time-keeping mechanism through embodied interactions but later internalized it, enabling it to perform the timing task with different environmental inputs. This process parallels the evolution of biological mechanisms such as circadian rhythms, where an organism builds an internal time representation through interactions with the environment and ultimately uses it as a buffer to adapt to environmental changes

The findings from this study, in conjunction with those of \cite{deverett2019interval, lin2023temporal}, suggest several implications for interval timing tasks. Specifically, the temporal representation required for successfully completing these tasks appears to be extractable by an artificial neural network from its input. This could imply the involvement of lower-level cognitive processes \citep{konig2013unifying} in the execution of these tasks. This observation further supports the distinction made in time perception between duration judgment (i.e., the ability to estimate or reproduce time intervals) and the passage of time (i.e., how time feels) \citep{thones2019standard}. Specifically, while an artificial agent can learn to perform interval timing tasks with high accuracy,  evaluating the subjective passage of time (whether time feels fast or slow), remains a significantly more challenging problem for such systems. This disparity may indicate the differing levels of cognitive complexity involved in the two tasks. 

This study has several limitations. The parallels drawn between the agent’s behavior and biological systems, namely the resemblance to the SBF model of timing and the evolution of circadian rhythms, are preliminary and require further investigation. Given the complexity of biological systems, such analogies may overlook critical nuances. Two distinct systems arriving at similar solutions may not always imply shared mechanisms. This is because, the same problem can often be solved through multiple, distinct optimization pathways. Additionally, our analysis focused solely on the action network, whereas previous work has examined the value network and reported ramping-cell-like activity for temporal tasks \citep{lin2023temporal}. Future work could extend this study by analyzing the value network in our agent to uncover additional biologically plausible temporal patterns in the artificial neurons. Finally, this work represents an initial step towards using artificial agents to study temporal processing in biological systems. Understanding the implications of these similarities and translating them into meaningful biological insights remains a long-term challenge.

\section{Appendix}

\vspace{-5pt}
\begin{figure}[H]
  % Row 1
    \begin{subfigure}[b]{0.5\textwidth}
        \centering
        \includegraphics[width=\textwidth, height=3cm]{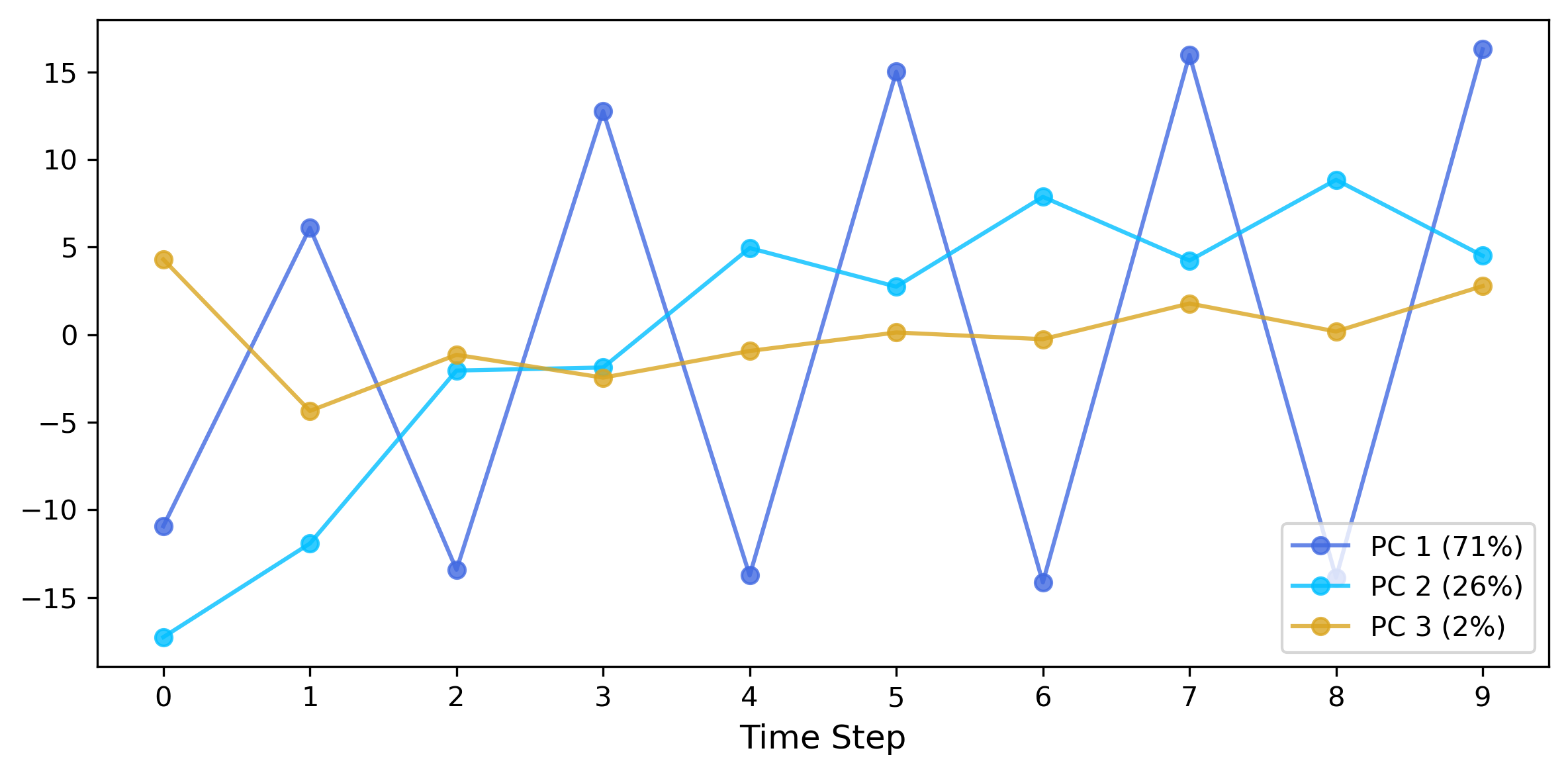}
        \caption{Target duration: 2 time steps}
    \end{subfigure}
    \hfill
    \begin{subfigure}[b]{0.5\textwidth}
        \centering
        \includegraphics[width=\textwidth, height=3cm]{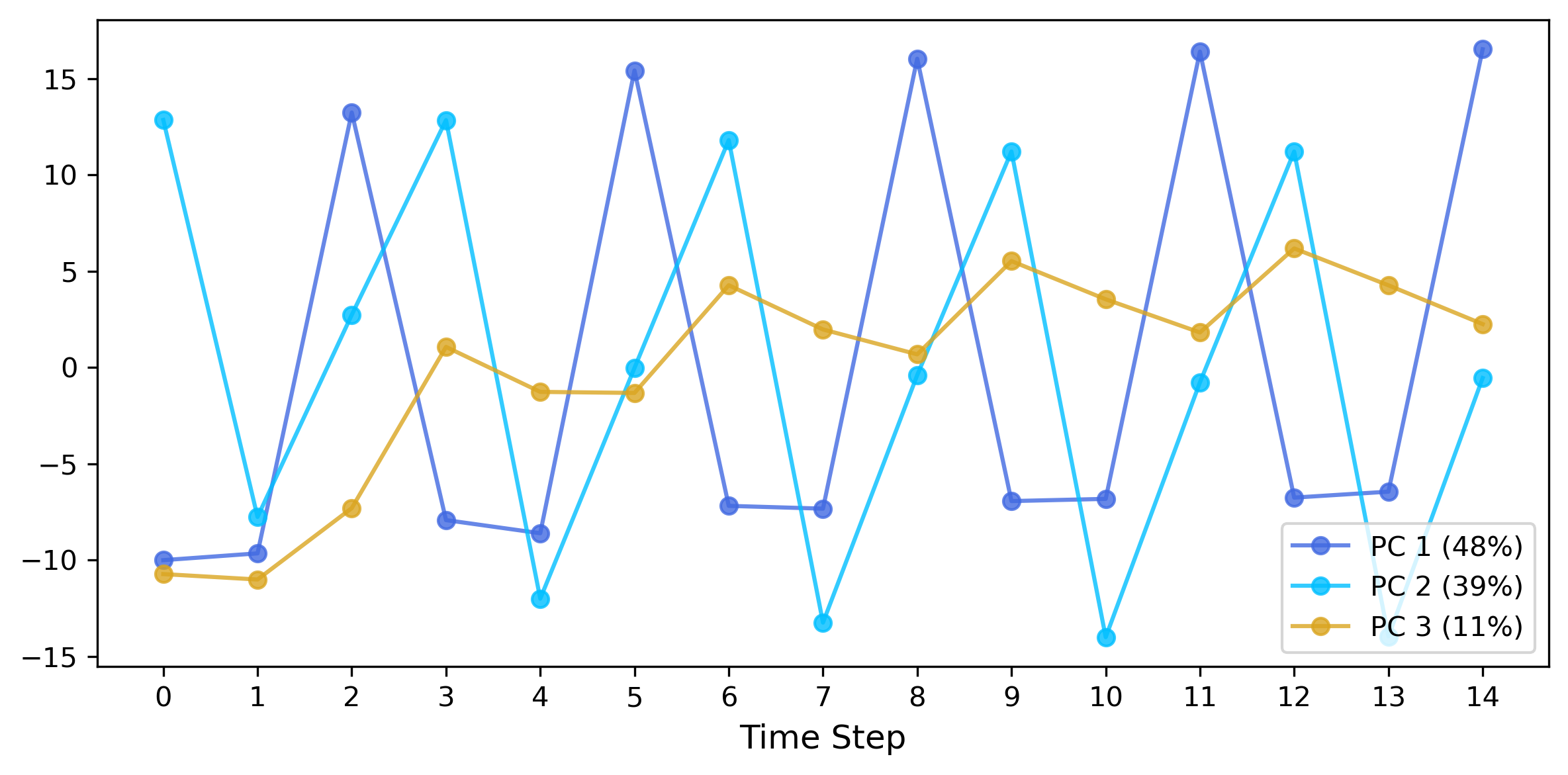}
        \caption{Target duration: 3 time steps}
    \end{subfigure}
    \hfill
    \begin{subfigure}[b]{0.5\textwidth}
        \centering
        \includegraphics[width=\textwidth,,height=3cm]{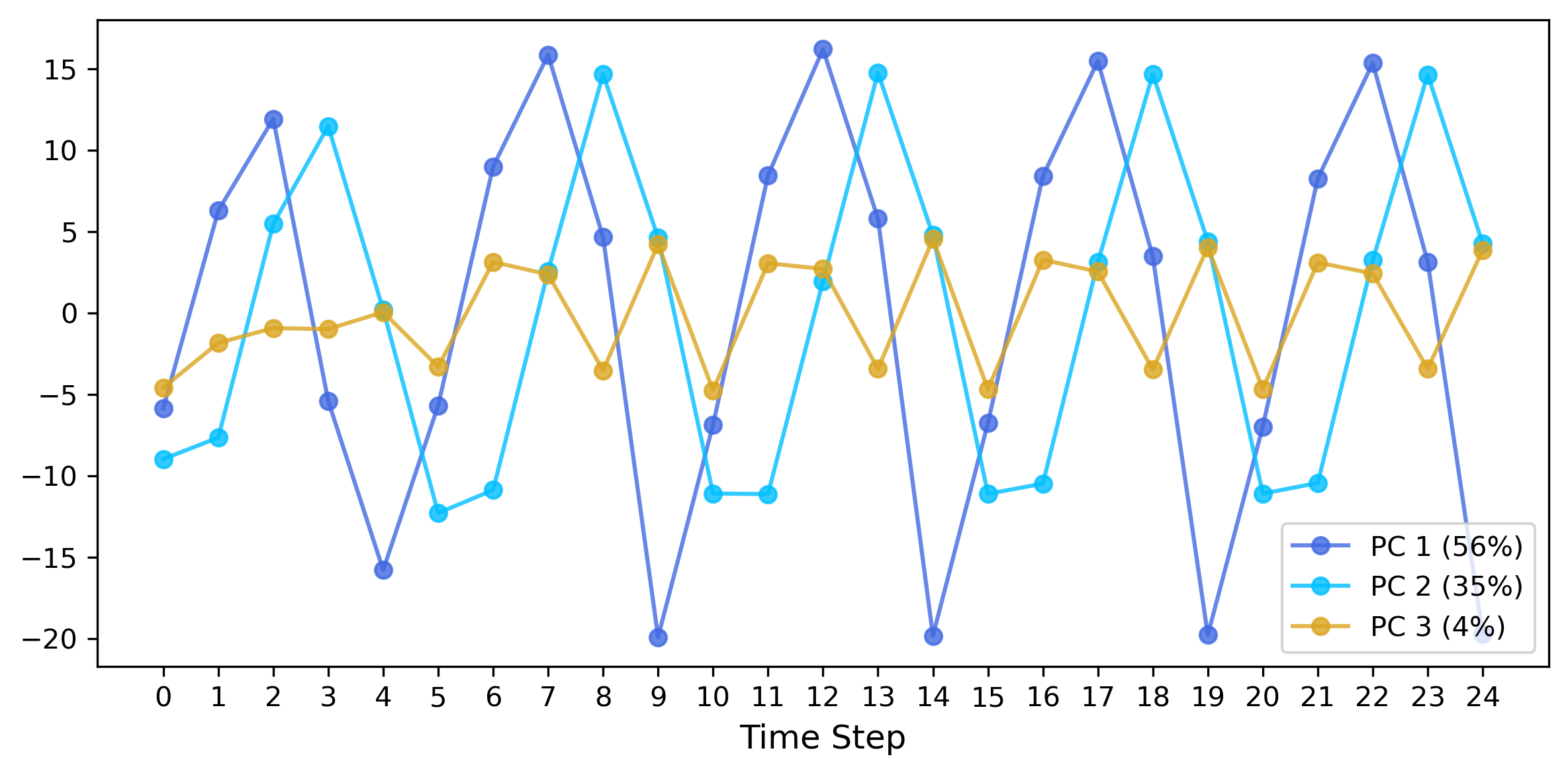}
        \caption{Target duration: 5 time steps}
    \end{subfigure}
    \caption{PCA of the 256 LSTM hidden state neurons across time for the time production task with different target durations}
     \label{fig:appendix_pca}
\end{figure}

\begin{figure}[H]
 
    \begin{subfigure}[b]{0.5\textwidth}
        \centering
        \includegraphics[width=\textwidth, height=2.5cm, keepaspectratio]{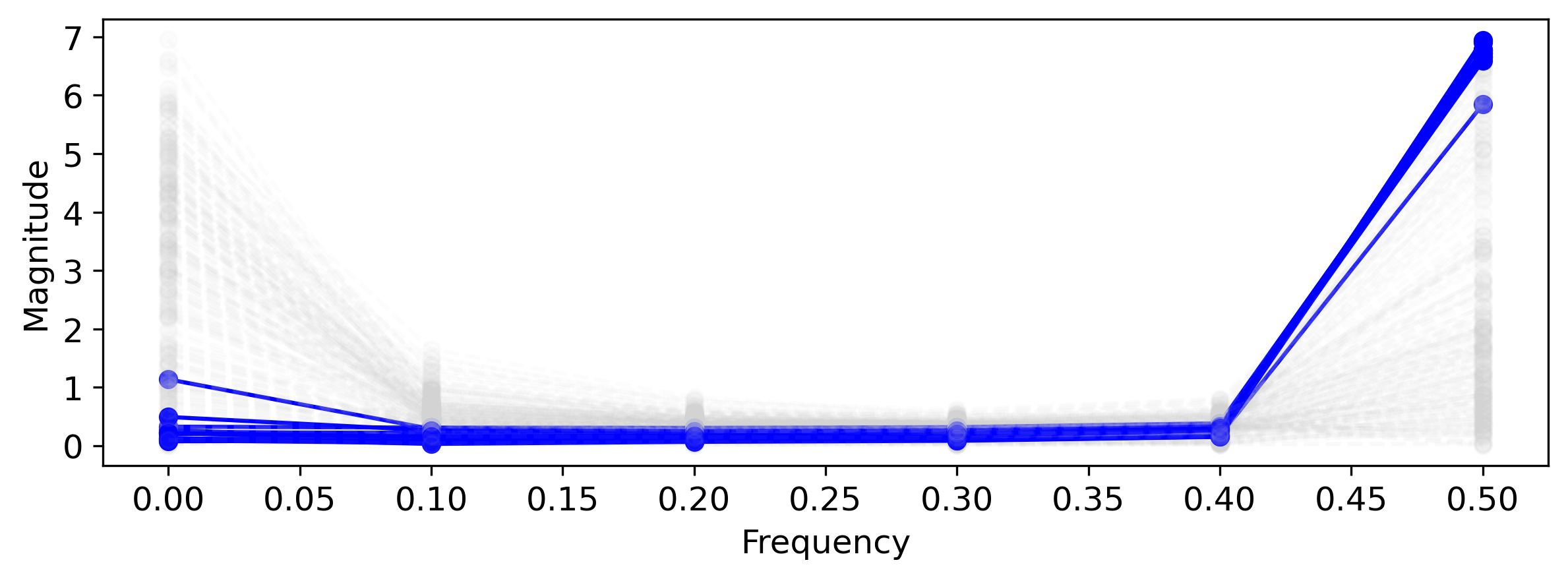}
        \caption{Target duration: 2 time steps}
    \end{subfigure}
    % \hfill
    \begin{subfigure}[b]{0.5\textwidth}
        \centering
        \includegraphics[width=\textwidth, height=2.5cm, keepaspectratio]{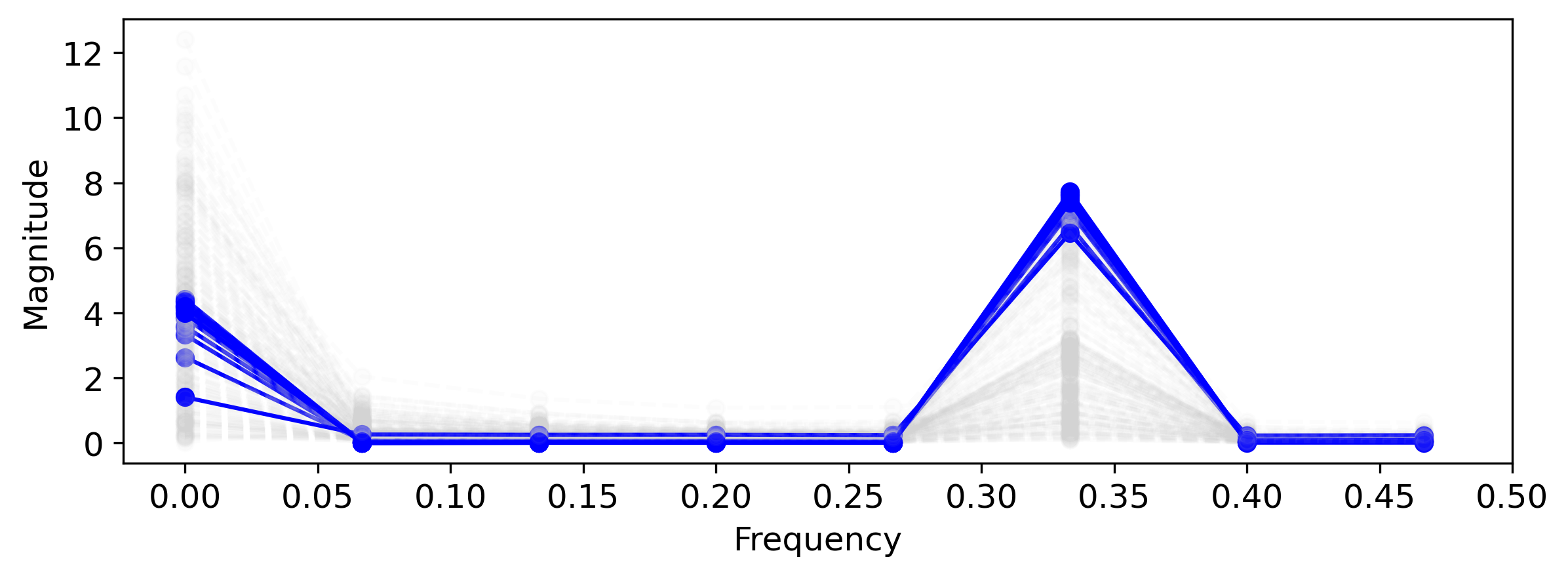}
        \caption{Target duration: 3 time steps}
    \end{subfigure}
    % \hfill
    \begin{subfigure}[b]{0.5\textwidth}
        \centering
        \includegraphics[width=\textwidth, height=2.5cm, keepaspectratio]{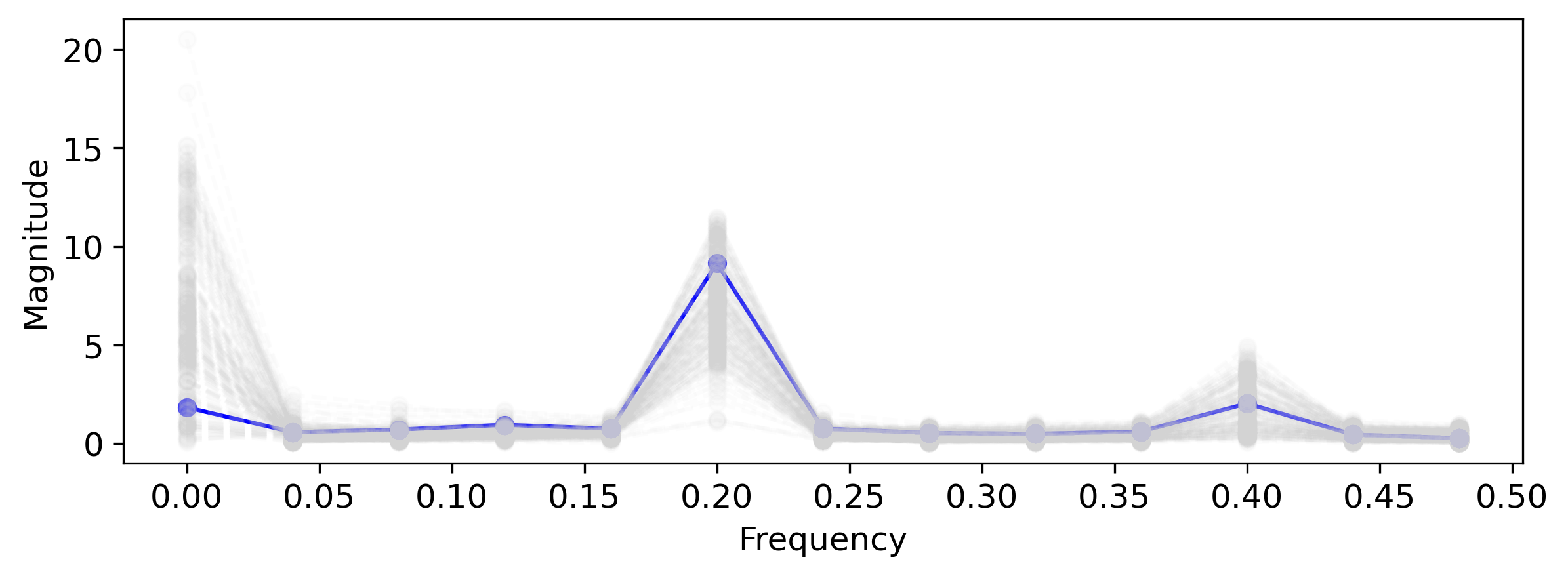}
        \caption{Target duration: 5 time steps}
    \end{subfigure}

    \caption{FFT of the 256 LSTM hidden state neurons across time for the time production task with different target durations}
     \label{fig:appendix_fft}

\end{figure}
\begin{figure}[H]

    % Row 3
    \begin{subfigure}[b]{0.5\textwidth}
        \centering
        \includegraphics[width=\textwidth]{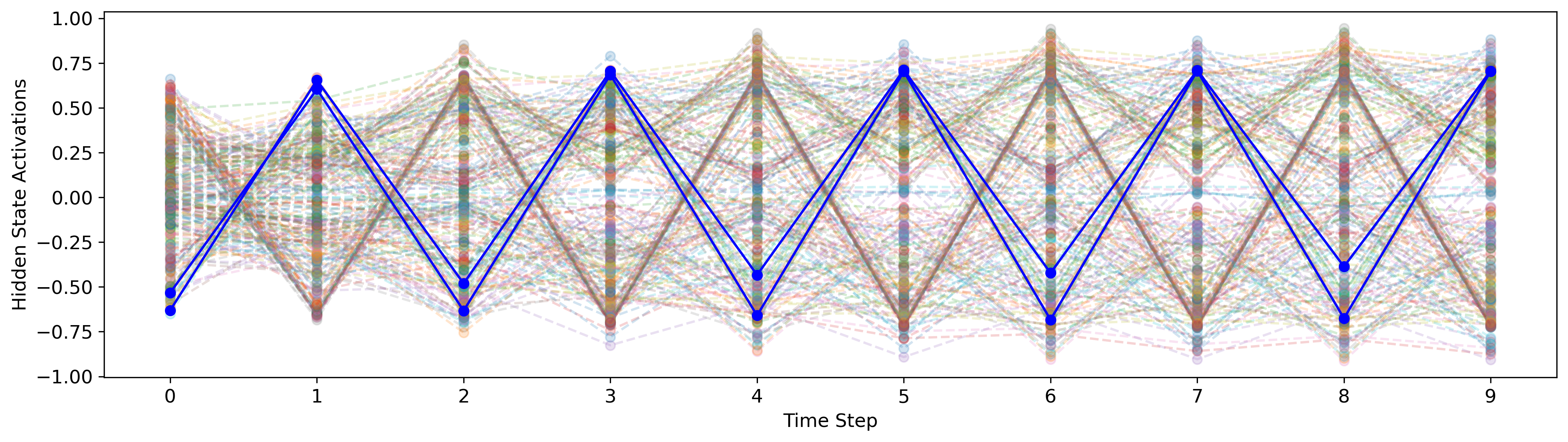}
        \caption{Target duration: 2 time steps}
    \end{subfigure}
    % \hfill
    \begin{subfigure}[b]{0.5\textwidth}
        \centering
        \includegraphics[width=\textwidth]{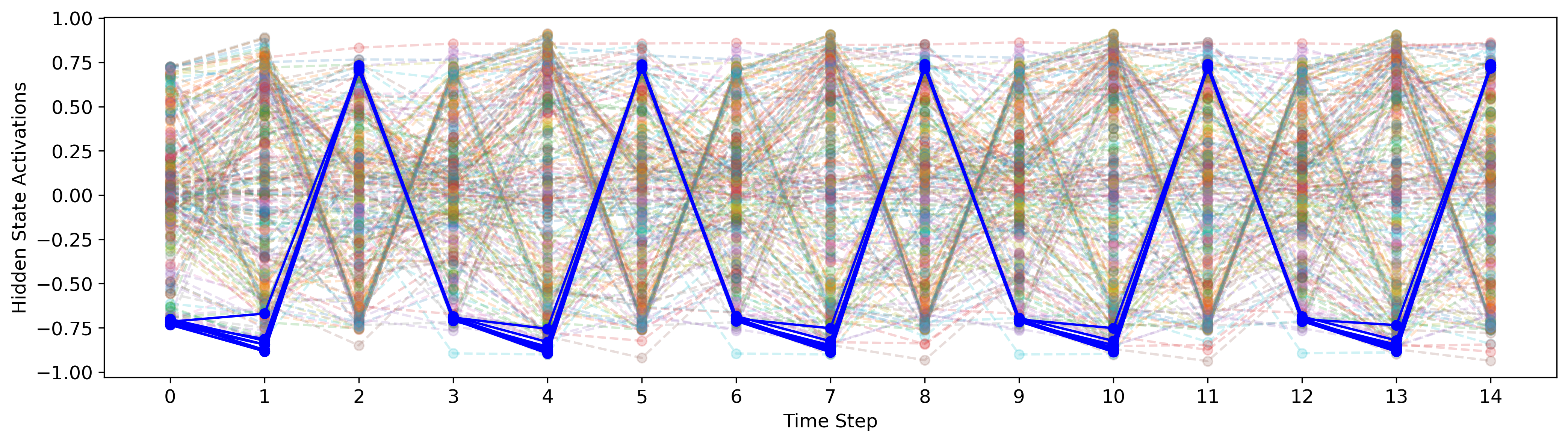}
        \caption{Target duration: 3 time steps}
    \end{subfigure}
    % \hfill
    \begin{subfigure}[b]{0.5\textwidth}
        \centering
        \includegraphics[width=\textwidth]{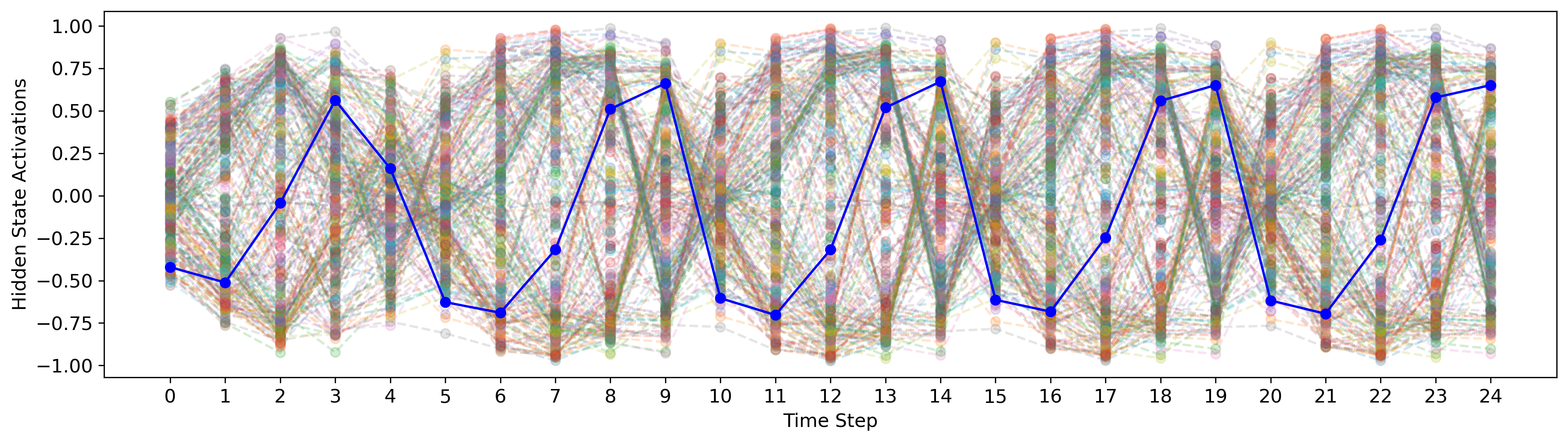}
        \caption{Target duration: 5 time steps}
    \end{subfigure}

    \caption{Raw activations of the 256 LSTM hidden state neurons across time for the time production task with different target durations}
    \label{fig:appendix_raw_activations}

\end{figure}

% The DRL agent for the time production task was trained for different target durations, namely, 2,3,4,5 time steps. The PCA and FFT of the LSTM hidden state activations show consistent findings across time steps. These findings are as follows, oscillation is a predominant patterns in the LSTM hidden state activations as exhibited by PCA of the activations across time. Additionally, these are high amplitude oscillations with frequencies corresponding to the target interval as exhibited by FFT analysis of the activations. Figure x shows the PCA and FFT results for duration 2, 3 and 5. 

 \clearpage

\section{Acknowledgements}
This work was supported by European Union’s Horizon 2020 FET research program under grant agreement
No. 964464 (ChronoPilot).

\footnotesize
\bibliographystyle{apalike}
\bibliography{references} % replace by the name of your .bib file

@article{deverett2019interval,
  title={Interval timing in deep reinforcement learning agents},
  author={Deverett, Ben and Faulkner, Ryan and Fortunato, Meire and Wayne, Gregory and Leibo, Joel Z},
  journal={Advances in Neural Information Processing Systems},
  volume={32},
  year={2019}
}

@article{maniadakis2015integrated,
  title={Integrated intrinsic and dedicated representations of time: a computational study involving robotic agents},
  author={Maniadakis, Michail and Trahanias, Panos},
  journal={Timing \& Time Perception},
  volume={3},
  number={3-4},
  pages={246--268},
  year={2015},
  publisher={Brill}
}

@article{hochreiter1997long,
  title={Long short-term memory},
  author={Hochreiter, Sepp and Schmidhuber, J{\"u}rgen},
  journal={Neural computation},
  volume={9},
  number={8},
  pages={1735--1780},
  year={1997},
  publisher={MIT press}
}

@article{yu2019review,
  title={A review of recurrent neural networks: LSTM cells and network architectures},
  author={Yu, Yong and Si, Xiaosheng and Hu, Changhua and Zhang, Jianxun},
  journal={Neural computation},
  volume={31},
  number={7},
  pages={1235--1270},
  year={2019},
  publisher={MIT Press One Rogers Street, Cambridge, MA 02142-1209, USA journals-info~…}
}

@article{matell2000neuropsychological,
  title={Neuropsychological mechanisms of interval timing behavior},
  author={Matell, Matthew S and Meck, Warren H},
  journal={Bioessays},
  volume={22},
  number={1},
  pages={94--103},
  year={2000},
  publisher={Wiley Online Library}
}

@article{buhusi2005makes,
  title={What makes us tick? Functional and neural mechanisms of interval timing},
  author={Buhusi, Catalin V and Meck, Warren H},
  journal={Nature reviews neuroscience},
  volume={6},
  number={10},
  pages={755--765},
  year={2005},
  publisher={Nature Publishing Group UK London}
}

@article{heltberg2021tale,
  title={A tale of two rhythms: Locked clocks and chaos in biology},
  author={Heltberg, Mathias L and Krishna, Sandeep and Kadanoff, Leo P and Jensen, Mogens H},
  journal={Cell Systems},
  volume={12},
  number={4},
  pages={291--303},
  year={2021},
  publisher={Elsevier}
}

@article{hut2011evolution,
  title={Evolution of time-keeping mechanisms: early emergence and adaptation to photoperiod},
  author={Hut, Roelof A and Beersma, Domien GM},
  journal={Philosophical Transactions of the Royal Society B: Biological Sciences},
  volume={366},
  number={1574},
  pages={2141--2154},
  year={2011},
  publisher={The Royal Society}
}

@article{seki2022evolution,
  title={Evolution of self-sustained circadian rhythms is facilitated by seasonal change of daylight},
  author={Seki, Motohide and Ito, Hiroshi},
  journal={Proceedings of the Royal Society B},
  volume={289},
  number={1987},
  pages={20220577},
  year={2022},
  publisher={The Royal Society}
}

@article{buzsaki2013scaling,
  title={Scaling brain size, keeping timing: evolutionary preservation of brain rhythms},
  author={Buzs{\'a}ki, Gy{\"o}rgy and Logothetis, Nikos and Singer, Wolf},
  journal={Neuron},
  volume={80},
  number={3},
  pages={751--764},
  year={2013},
  publisher={Elsevier}
}

@article{kanwisher2023using,
  title={Using artificial neural networks to ask ‘why’questions of minds and brains},
  author={Kanwisher, Nancy and Khosla, Meenakshi and Dobs, Katharina},
  journal={Trends in Neurosciences},
  volume={46},
  number={3},
  pages={240--254},
  year={2023},
  publisher={Elsevier}
}

@article{richards2019deep,
  title={A deep learning framework for neuroscience},
  author={Richards, Blake A and Lillicrap, Timothy P and Beaudoin, Philippe and Bengio, Yoshua and Bogacz, Rafal and Christensen, Amelia and Clopath, Claudia and Costa, Rui Ponte and de Berker, Archy and Ganguli, Surya and others},
  journal={Nature neuroscience},
  volume={22},
  number={11},
  pages={1761--1770},
  year={2019},
  publisher={Nature Publishing Group US New York}
}

@article{lin2023temporal,
  title={Temporal encoding in deep reinforcement learning agents},
  author={Lin, Dongyan and Huang, Ann Zixiang and Richards, Blake Aaron},
  journal={Scientific Reports},
  volume={13},
  number={1},
  pages={22335},
  year={2023},
  publisher={Nature Publishing Group UK London}
}

@article{thones2019standard,
  title={A standard conceptual framework for the study of subjective time},
  author={Th{\"o}nes, Sven and Stocker, Kurt},
  journal={Consciousness and cognition},
  volume={71},
  pages={114--122},
  year={2019},
  publisher={Elsevier}
}

@article{wittmann2009inner,
  title={The inner experience of time},
  author={Wittmann, Marc},
  journal={Philosophical Transactions of the Royal Society B: Biological Sciences},
  volume={364},
  number={1525},
  pages={1955--1967},
  year={2009},
  publisher={The Royal Society London}
}

@article{salih2020dynamic,
  title={Dynamic scene change detection in video coding},
  author={Salih, Y and George, LE},
  journal={International Journal of Engineering},
  volume={33},
  number={5},
  pages={966--974},
  year={2020}
}

@article{towers2024gymnasium,
  title={Gymnasium: A standard interface for reinforcement learning environments},
  author={Towers, Mark and Kwiatkowski, Ariel and Terry, Jordan and Balis, John U and De Cola, Gianluca and Deleu, Tristan and Goulao, Manuel and Kallinteris, Andreas and Krimmel, Markus and KG, Arjun and others},
  journal={arXiv preprint arXiv:2407.17032},
  year={2024}
}

@article{stable-baselines3,
  author  = {Antonin Raffin and Ashley Hill and Adam Gleave and Anssi Kanervisto and Maximilian Ernestus and Noah Dormann},
  title   = {Stable-Baselines3: Reliable Reinforcement Learning Implementations},
  journal = {Journal of Machine Learning Research},
  year    = {2021},
  volume  = {22},
  number  = {268},
  pages   = {1-8},
  url     = {http://jmlr.org/papers/v22/20-1364.html}
}

@article{konig2013unifying,
  title={A unifying approach to high-and low-level cognition},
  author={K{\"o}nig, Peter and K{\"u}hnberger, Kai-Uwe and Kietzmann, Tim C},
  journal={Models, simulations, and the reduction of complexity},
  volume={4},
  pages={117--139},
  year={2013},
  publisher={De Gruyter Berlin, Germany}
}

@article{yu2017continuous,
  title={Continuous timescale long-short term memory neural network for human intent understanding},
  author={Yu, Zhibin and Moirangthem, Dennis S and Lee, Minho},
  journal={Frontiers in neurorobotics},
  volume={11},
  pages={42},
  year={2017},
  publisher={Frontiers Media SA}
}

@article{beer1995dynamics,
  title={On the dynamics of small continuous-time recurrent neural networks},
  author={Beer, Randall D},
  journal={Adaptive Behavior},
  volume={3},
  number={4},
  pages={469--509},
  year={1995},
  publisher={Sage Publications Sage CA: Thousand Oaks, CA}
}

\end{document}